\documentclass[aps,prd, groupedaddress,amssymb,preprint,showpacs,epsfig,nofootinbib]{revtex4}

\usepackage{graphicx}
\usepackage{bm}
\usepackage{dcolumn}

\usepackage{amsmath}
\usepackage{amssymb}

\usepackage{epsfig}

\begin{document}
\newcommand{\newc}{\newcommand}

\newc{\be}{\begin{equation}}
\newc{\ee}{\end{equation}}
\newc{\ba}{\begin{eqnarray}}
\newc{\ea}{\end{eqnarray}}
\newc{\bea}{\begin{eqnarray*}}
\newc{\eea}{\end{eqnarray*}}
\newc{\D}{\partial}
\newc{\ie}{{\it i.e.} }
\newc{\eg}{{\it e.g.} }
\newc{\etc}{{\it etc.} }
\newc{\etal}{{\it et al.}}
\newc{\lcdm}{$\Lambda$CDM}
\newcommand{\nn}{\nonumber}
\newc{\ra}{\rightarrow}
\newc{\lra}{\leftrightarrow}
\newc{\lsim}{\buildrel{<}\over{\sim}}
\newc{\gsim}{\buildrel{>}\over{\sim}}
\newcommand{\mincir}{\raise
-3.truept\hbox{\rlap{\hbox{$\sim$}}\raise4.truept\hbox{$<$}\ }}
\newcommand{\magcir}{\raise
-3.truept\hbox{\rlap{\hbox{$\sim$}}\raise4.truept\hbox{$>$}\ }}

\title{Massive gravitons dark matter scenario revisited}

\author{Hyung Won Lee}
\email{hwlee@inje.ac.kr}

\author{Kyoung Yee Kim}
\email{kimky@inje.ac.kr}

\author{Yun Soo Myung}
\email{ysmyung@inje.ac.kr}

\affiliation{Institute of Basic Science and School of Computer Aided
Science, Inje University, Gimhae 621-749, Korea}

\begin{abstract}
We reexamine the massive graviton dark matter scenario (MGCDM) which
was recently considered as an alternative to dark energy models.
When introducing the native and effective equations of state (EoS),
it is shown that there is no phantom phase in the evolution toward
the far past. Also we show that  the past accelerating phase arises
from the interaction between massive graviton and cold dark matter.

\end{abstract}
\maketitle

\section{Introduction}
Recently the massive graviton dark matter scenario (MGCDM)  which
has been originally proposed by Visser~\cite{Vis} was considered as
an alternative dark energy model to explaining the present
accelerating universe~\cite{AMA,BPAL}. In order to explain the
evolution of the universe, they used the ``geometric" dark energy
EoS mainly as~\cite{BPAL}
\begin{equation}
\label{eos1} w_{\rm DE}(a)=-1-\frac{1}{3}\;\frac{d{\rm ln}\delta
H^{2}}{d{\rm ln}a}=-  1 - \frac{2}{3} \left(\frac{7 - 10a^2}{7 -
5a^2}\right),
\end{equation}
which shows that the massive graviton theory  can be treated as an
additional effective fluid with EoS (\ref{eos1}) cosmologically.
Also they have emphasized  that the current Hubble function has only
two free parameters  of $H_0$ and $\Omega_{m}$ which are  the same
number of parameters as the $\Lambda$CDM model. However,  this MGCDM
showed a different phantom phase of $\lim_{a\to0}w_{\rm DE}(a)=-5/2$
and a current decelerating phase~\cite{BPAL}, in compared to the
$\Lambda$CDM model with $w_{\rm \Lambda}=-1$. This is mainly because
they used the geometric dark energy EoS $w_{\rm DE}(a)$.

On the other hand, it is well known that if the conservation law is
modified due to the presence of other matter as
(\ref{conservation}), one has to introduce the effective EoS
parameter $w_g^{\rm eff}(a)$ as well as the native EoS $w_g^{\rm
nat}(a)$ to describe the evolution of the universe
correctly~\cite{ZP1}. For the holographic dark energy model, two of
authors have clarified that although there is a phantom phase when
using the native EoS~\cite{WGA}, there is no phantom phase when
using the effective EoS~\cite{KLM,KLMb}.

In this work, we wish to reexamine the evolution of the universe
based on the MGCDM by introducing the native and effective EoS. It
is hard to derive any phantom phase when using the effective EoS
$w_g^{\rm eff}$ instead of the geometric EoS $w_{\rm DE}$. Finally,
we show that the past accelerating phase arises from the interaction
between massive graviton and CDM.

\section{\label{sec:two} Massive graviton}

%%%%%%%%%%%%%%%%%%%%%%%%%%%%%%%%%%%%%%%%%%%%%%%%%%%%%%%%%%%%%%%%%%%%%%%%%%%%%%%%
We briefly review the  Visser's massive gravity approach~\cite{Vis}.
The action is given by
\begin{eqnarray}
\label{fullaction} S=\int d^4x\left[\sqrt{-g}\frac{c^4 R(g)}{16\pi
G} + {\cal{L}}_{\rm mass}(g,g_0) +{\cal{L}}_{\rm matter}(g)\right]
\end{eqnarray}
where the first term is  the Einstein-Hilbert Lagrangian and the
last is the Lagrangian of matter. The second term is designed for
the massive graviton expressed in terms of the  bi-metric $(g_0,g)$
as
\begin{eqnarray}
{\cal{L}}_{\rm mass}(g,g_0) = \frac{1}{2}\frac{{m_g}^2 c^2}{\hbar^2}
\sqrt{-g_0}\bigg\{ ( g_0^{-1})^{\mu\nu} ( g-g_0)_{\mu\sigma}(
g_0^{-1})^{\sigma\rho}
\\ \nonumber
 \times ( g-g_0)_{\rho\nu}-\frac{1}{2}
\left[( g_0^{-1})^{\mu\nu}( g-g_0)_{\mu\nu}\right]^2\bigg\},
\end{eqnarray}
where $m_g$ is the graviton mass and
$(g_0)_{\mu\nu}$ is a general flat metric.

The Einstein equations  takes the form
\begin{equation}\label{field-equations}
G^{\mu\nu} -\frac{1}{2}\frac{{m_g}^2 c^2}{\hbar^2} M^{\mu\nu} = -\frac{8\pi G}{c^4}  T^{\mu\nu},
\end{equation}
where $G^{\mu\nu}$ is the Einstein tensor, $T^{\mu\nu}$ is the
energy-momentum tensor for a perfect fluid, and
\begin{eqnarray}\label{massive tensor}
M^{\mu\nu} =  (g_0^{-1})^{\mu\sigma}\bigg[ (g-g_0)_{\sigma\rho} -
\frac{1}{2}(g_0)_{\sigma\rho}(g_0^{-1})^{\alpha\beta}\times(g-g_0)_{\alpha\beta}
\bigg](g_0^{-1})^{\rho\nu}.
\end{eqnarray}
In the limit $m_g\rightarrow 0$, we recover the Einstein  equation
is recovered. Importantly, since the Einstein tensor should satisfy
the Bianchi identity $\nabla_\nu G^{\mu\nu} = 0$, the
non-conservation law of  the energy-momentum  tensor is  obtained as
 \begin{equation}\label{conservation}
   \nabla_\nu T^{\mu\nu} = \frac{{m_g}^2 c^6}{16\pi G \hbar^2 } \nabla_\nu M^{\mu\nu}
 \end{equation}
which will play the crucial role in the cosmological evolution.

\section{\label{evolution}Cosmological evolution}
In order to apply the action (\ref{fullaction})  to cosmology, we
need to introduce the bi-metric explicitly. First we use the
Friedmann-Roberston-Walker metric as the dynamical one
\begin{equation}\label{dyn-metric}
ds^2_{\rm FRW} = c^2 dt^2 - a^2(t) \left [ dr^2 + r^2 d\Omega_2^2
\right ]
\end{equation}
with $a(t)$  the scale factor. The flat  metric expressed in terms
of spherical coordinates is proposed to be the static one:
\begin{equation}\label{flat-metric}
ds_0^2 = c^2 dt^2 - \left [ dr^2 + r^2 d\Omega_2^2 \right ].
\end{equation}

The evolution  of a MGCDM cosmology is governed  by two Friedmann
equations:
 \begin{equation}\label{eqfried1}
 \left( \frac{\dot{a}}{a}\right)^2 =  \frac{8\pi G}{3 c^2} \rho
             + \frac{{m_g}^2 c^4 }{4 \hbar^2 }(1 - a^2),
\end{equation}
 \begin{equation}\label{eqfried2}
  \frac{\ddot{a}}{a} + \frac{1}{2}\left( \frac{\dot{a}}{a}\right)^2 =
               - \frac{4\pi G}{c^2} p - \frac{{m_g}^2 c^4 }{8 \hbar^2 }a^2(a^2-1),
 \end{equation}
where $\rho$ is the energy density and  $p$ is the pressure. From
the observation of  Eqs. (\ref{eqfried1}) and (\ref{eqfried2}), one
can read off the  energy density and pressure for massive graviton
as
\begin{equation} \label{energy-massive}
\rho_g = \frac{3 m_g^2 c^6 }{32 \pi G \hbar^2} (1-a^2),
\end{equation}
\begin{equation} \label{pressure-massive}
p_g = \frac{m_g^2 c^6 }{32 \pi G \hbar^2} a^2 (a^2-1).
\end{equation}
We note that the positive energy density and negative pressure is
allowed to explain the accelerating universe. In this case, it
requires
\begin{equation} \label{ineq}
a^2 <1.
\end{equation}
Hence the native equation of state is simply given by
\begin{equation}\label{bare-eos}
w_g^{\rm nat}(a) \equiv \frac{p_g}{\rho_g}=-\frac{1}{3} a^2,
\end{equation}
whose limit of $a \to 0$ is zero.  From  (\ref{conservation}),  we
find the the non-conservation law for the matter including cold dark
matter (CDM)
 \begin{equation}\label{conserve-matter}
   \dot{\rho} + 3 H (\rho + p)  =
     - 3 H \frac{{m_g}^2 c^6 }{32\pi G \hbar^2} (a^4 - 6a^2 +
     3)\equiv -3H Q
 \end{equation}
with  $H = \dot{a}/a$ the Hubble parameter and
 \begin{equation}\label{Qvalue}
   Q = \frac{{m_g}^2 c^6 }{32\pi G \hbar^2} (a^4 - 6a^2 +
     3).
 \end{equation}
   On the other hand, one can obtain the non-conservation law  for
massive graviton  by computing (\ref{energy-massive}) and
(\ref{pressure-massive}) directly
 \begin{equation}\label{conserve-graviton}
   \dot{\rho_g} + 3 H (\rho_g + p_g)  =
      3 H Q.
 \end{equation}
Rewriting  Eq. (\ref{conserve-graviton}) as
\begin{equation}\label{conserve-graviton-eff}
   \dot{\rho_g} + 3 H (1 + w_g^{\rm eff} ) \rho_g  = 0,
\end{equation}
we can define  the effective equation of state for massive
graviton~\cite{ZP1}
\begin{equation}\label{eos-eff-graviton}
w^{\rm eff}_g(a) = - \frac{1-\frac{5}{3}a^2}{1-a^2}
\end{equation}
whose limit of $a\to 0$ is $-1$.  When introducing the massive
graviton ${\cal L}_{\rm mass}$, both matter and massive graviton do
not satisfy their own conservation law as is shown by
(\ref{conserve-matter}) and (\ref{conserve-graviton}).  Instead, the
total mixture of matter and massive graviton fluid satisfies the
conservation law as
\begin{equation}\label{conserve-total}
{\dot\rho}_t + 3 H (\rho_t + p_t ) = 0,
\end{equation}
with $\rho_t = \rho + \rho_g$ and $p_t = p + p_g$. This shows
clearly that our picture is quite different from (\ref{eos1})
imposed by  Ref.\cite{AMA,BPAL}. In other words, in order to express
the evolution of universe due to the massive graviton properly, we
will use its own native EoS $w_{g}^{\rm nat}$ and effective EoS
$w_{g}^{\rm eff}$ instead of  $w_{\rm DE}$. This is a well-accepted
approach to cosmology when two different matters coexist in the
universe, showing that either  decaying of massive graviton to CDM
for $Q<0$ or decaying of CDM  to massive graviton for $Q>0$ as in
quintessence~\cite{ZP1},  $\Lambda(t)$CDM model~\cite{WM}, and the
Brans-Dicke cosmology~\cite{LKM}.

Now we are in a position to solve Eq. (\ref{eqfried1}) for $a(t)$.
For this purpose, Eq. (\ref{eqfried1}) can be written as
\begin{equation}\label{hubble}
H^2 = H_0^2 \left [ \chi \Omega_m^0 \left ( \frac{a_0}{a} \right )^4 +
  \Omega_m^0 \left ( \frac{a_0}{a} \right )^3 +
  \frac{\alpha_g^2}{4} \left (1-a^2 \right )\right ],
\end{equation}
where $H_0$ is the Hubble constant at current, $\chi$ is the current
ratio of radiation density to dark matter density, and $\Omega_m^0$
is the current density parameter of CDM.  $\alpha_g$ is the ratio of
graviton mass and Hubble mass
\begin{equation}\label{mass-ratio}
\alpha_g = \frac{m_g}{m_H}, \,\,\, m_H = \frac{\hbar H_0}{c^2} =
3.8026\times 10^{-69} h_0 {\rm kg},
\end{equation}
where $h_0$ is defined as $H_0 = 100h_0$ km/s/Mpc. Evaluating Eq.
(\ref{hubble}) at the present time, the current scale factor is
determined to be
\begin{equation}\label{a0}
a_0 = \sqrt{1- \frac{4}{\alpha_g^2}\Big(1 - \Omega_m^0 - \chi
\Omega_m^0 \Big)}.
\end{equation}
In order to have real value for $a_0$,  one requires
\begin{equation}\label{low-limit-alpha}
\alpha_g > 2 \sqrt{1-\Omega_m^0 - \chi \Omega_m^0},
\end{equation}
which  corresponds to the inequality on the mass of massive graviton
as
\begin{equation}\label{mass-low}
m_g > 2 \sqrt{1-\Omega_m^0 - \chi \Omega_m^0} m_H .
\end{equation}
Taking $\Omega_m^0=0.27$ and $\chi \simeq 3.1\times 10^{-4}$,  we
have two bounds on $\alpha_g$ and $m_g$
\begin{equation}\label{mass-low0}
\alpha_g > 1.71, \,\,\, m_g > 1.71 m_H \simeq 6.50 \times 10^{-69}
h_0 {\rm  kg} .
\end{equation}
$\alpha_g$ is depicted as a function of $\Omega_m^0$ in Fig. 1 and
its current value is $\alpha_g(\Omega_m^0)=1.71$.
\begin{figure}[ht]
\label{alphag} \mbox{\epsfxsize=8.5cm \epsffile{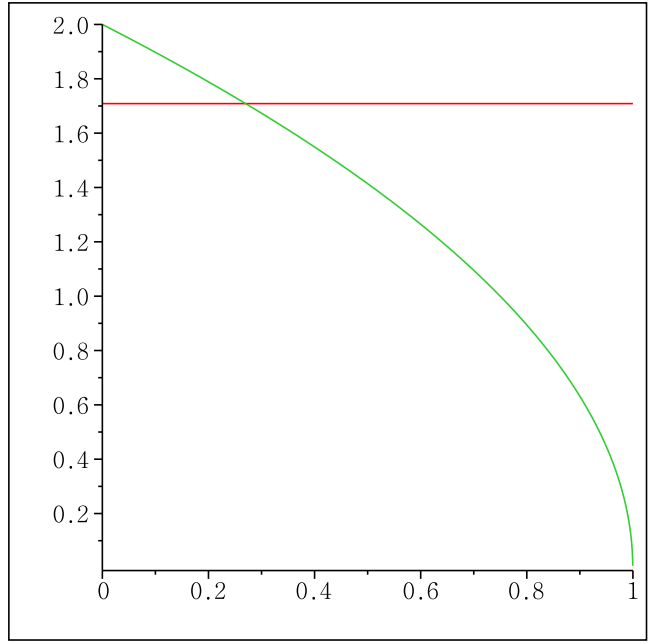}}
\caption{$\alpha_g$ as a function of current matter density
parameter, $\Omega_m^0$. Horizontal line represents  its current
value $\alpha_g(\Omega_m^0)=1.71$ at $\Omega_m^0 = 0.27$. }
\end{figure}

At this stage, we introduce  density parameters
\begin{equation}\label{omega}
\Omega_m = \frac{8 \pi G}{3H^2 c^2}\rho_m, \,\,\, \Omega_r = \frac{8 \pi G}{3H^2 c^2}\rho_r, \,\,\, \Omega_g = \frac{8 \pi G}{3H^2 c^2}\rho_g.
\end{equation}
Solving the Friedmann equations (\ref{hubble}) together with the
conservation laws (\ref{conserve-matter}) and
(\ref{conserve-graviton}), the relevant cosmological parameters are
determined to be
\begin{equation}\label{omegar}
\Omega_r = \frac{\chi \Omega_m^0 e^{-4x}}
       {\chi \Omega_m^0 e^{-4x} + \Omega_m^0 e^{-3x} + \frac{\alpha_g^2}{4} \left ( 1 - a_0^2 e^{2x}\right )},
\end{equation}
\begin{equation}\label{omegam}
\Omega_m = \frac{\Omega_m^0 e^{-3x}}
       {\chi \Omega_m^0 e^{-4x} + \Omega_m^0 e^{-3x} + \frac{\alpha_g^2}{4} \left ( 1 - a_0^2 e^{2x}\right )},
\end{equation}
\begin{equation}\label{omegag}
\Omega_g = \frac{\alpha_g^2}{4}\frac{1 - a_0^2 e^{2x}}
       {\chi \Omega_m^0 e^{-4x} + \Omega_m^0 e^{-3x} + \frac{\alpha_g^2}{4} \left ( 1 - a_0^2 e^{2x}\right )},
\end{equation}
\begin{equation}\label{h2}
H^2 = H_0^2 \left [
       {\chi \Omega_m^0 e^{-4x} + \Omega_m^0 e^{-3x} + \frac{\alpha_g^2}{4} \left ( 1 - a_0^2 e^{2x}\right )} \right ],
\end{equation}
\begin{equation}\label{wbare}
w_g^{\rm nat} = -\frac{a_0^2 e^{2x}}{3},
\end{equation}
\begin{equation}\label{weff}
w_g^{\rm eff} = -\frac{1-\frac{5}{3}a_0^2 e^{2x}}{1-a_0^2 e^{2x}}
\end{equation}
with a new variable  $x = \ln(a/a_0)\in[-\infty,\infty]$ including
$x=0$ at $a=a_0$, instead of scale factor $a$. The time evolution of
all parameters including density parameters, EoS, and Hubble
parameter is  shown as a function of $x$ in Fig. 2. This graphs
indicates that there is no phantom phase when using the native EoS
$w_g^{\rm nat}$ and the effective EoS $w_g^{\rm eff}$ but there is a
phantom phase when using the geometric  EoS $w_{\rm DE}$. This shows
clearly that the description  with $w_{\rm DE}$ is not appropriate
for interpreting  the modified  evolution due to the massive
graviton.
\begin{figure}[ht]
\label{evolve-fig} \mbox{\epsfxsize=8.5cm \epsffile{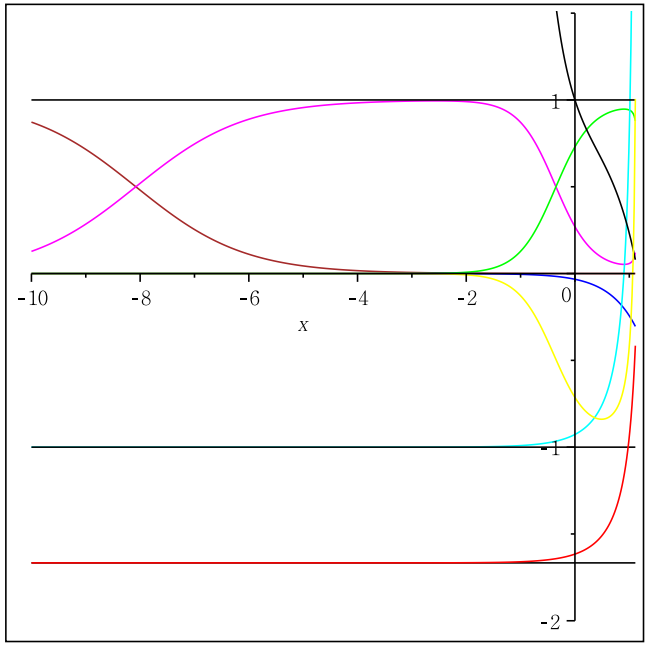}}
\caption{Cosmological evolution for there density parameters
[$\Omega_g$(green), $\Omega_m$(magenta), $\Omega_r$(brown)], four
equations of states [$w_g^{\rm nat}$(blue), $w_g^{\rm eff}$(cyan),
$w_{\rm DE}$(red), $w^{\rm eff}$(yellow\cite{AMA})], and Hubble parameter (black) as function of $x$,
for $\Omega_m^0 = 0.27$, $\chi=3.1\times10^{-4}$, and $\alpha_g =
1.8$. The bottom line corresponds to asymptote of $-5/3$ to $w_{\rm
DE}$ shown in Ref.\cite{BPAL}, while the second lowest line
represents asymptote of  $-1$ to $w_g^{\rm eff}$. }
\end{figure}

In order to compute the age of the universe,  we simply evaluate the
integration
\begin{equation}\label{age}
t_{U} = \int_0^{a_0} \frac{da}{a H(a)}.
\end{equation}
Substituting (\ref{hubble}) in (\ref{age}) leads to  the age of the
universe as shown Fig. 3 for chosen values of $\Omega_m^0$, showing
that the age of the universe is closely related to the mass of
massive graviton.  We note that $H_0^{-1} = 10.0 h_0^{-1}$Gy.
\begin{figure}[ht]
\label{age-fig} \mbox{\epsfxsize=8.5cm \epsffile{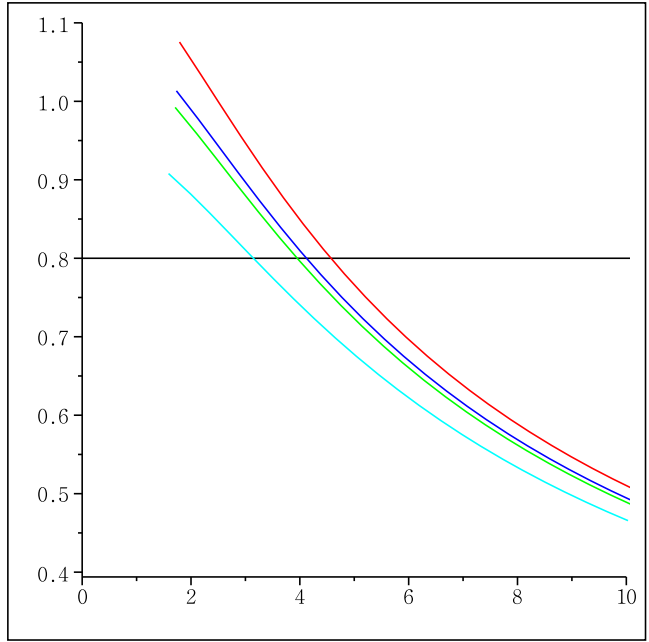}}
\caption{Relation between the age $t_U$ of the universe and the mass
ratio $\alpha_g$ for different values of $\Omega_m^0$, $0.20$(red),
$0.25$(blue), $0.27$(green), $0.37$(cyan), and horizontal line
denotes $0.8$, which corresponds to $8h_0^{-1}$ Gy. }
\end{figure}

At this stage, we would like to  mention that as is shown in Eq.
(\ref{hubble}), the Hubble parameter becomes zero at a certain value
of $a_c$ satisfying the condition of
\begin{equation}\label{critical-a}
\chi \Omega_m^0 \left ( \frac{a_0}{a_c} \right )^4 +
\Omega_m^0 \left ( \frac{a_0}{a_c} \right )^3 +
\frac{\alpha_g^2}{4} \left (1-a_c^2 \right ) = 0 .
\end{equation}
If one chooses $\alpha_g$ as the saturating bound  in Eq.
(\ref{low-limit-alpha}), $a_0$ becomes zero from Eq. (\ref{a0}) and
consequently, (\ref{critical-a}) implies that $a_c$ becomes  $1$.
However, this case  is meaningless because  the current scale factor
is chosen to be zero.  Hence $\alpha_g$ should satisfy the bound of
Eq. (\ref{low-limit-alpha}). This implies  that the mass of massive
graviton should be greater than a certain value to have a non-zero
scale factor ($m_g >1.71 m_H$)  at current time.  In this case, we
have the condition of $a_c
> 1$  which  means that the energy density of
massive graviton is negative as is shown in (\ref{energy-massive}).
In connection to this point, we may solve Eq. (\ref{hubble}) for $a$
as a function of cosmological time. The result  is depicted  in Fig.
4 for $\alpha_g=1.8$, $\Omega_m^0=0.27$, and
$\chi=3.1\times10^{-4}$, indicating that there is no sizable
difference when comparing to other cases.
\begin{figure}[ht]
\label{a-fig}
\mbox{\epsfxsize=8.5cm \epsffile{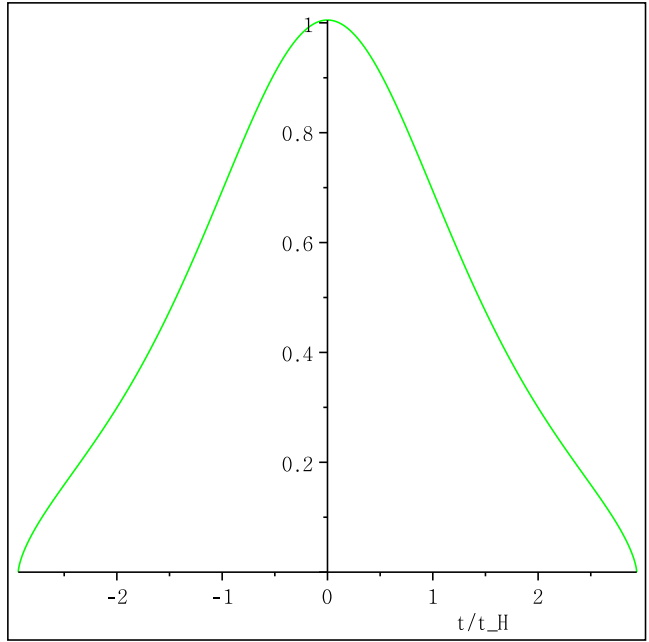}}
\caption{Time evolution for scale factor $a$ for
$\alpha_g=1.8$, $\Omega_m^0=0.27$, and $\chi=3.1\times10^{-4}$.
}
\end{figure}

Finally, we derive the geometric  equation of state $w_{\rm DE}$ in
(\ref{eos1}) from our approach.  We  start with the modified Hubble
equation
\begin{equation}\label{modified-hubble}
H^2 = H_0^2 \left [ \Omega_m^0 \left ( \frac{a_0}{a} \right )^3+
\delta H^2 \right ]
 = H_m^2 + H_\delta^2 ,
\end{equation}
where $ H_\delta^2$ includes the contribution from all matter
except CDM. Hence we can define `geometric' equations of state for
CDM and other matters including  massive graviton, respectively,
through
\begin{equation}\label{eff-eos-matter}
H_m^2 = \frac{8 \pi G}{3c^2}\rho_m^0 a_0^{3(1+w_m^{\rm geo})}
a^{-3(1+w_m^{\rm geo})} ,
\end{equation}
\begin{equation}\label{eff-eos-graviton}
H_\delta^2 = \frac{8 \pi G}{3c^2}\rho_g^0 a_0^{3(1+w_\delta^{\rm
geo})} a^{-3(1+w^{\rm geo}_\delta)}.
\end{equation}
Using $x = \ln(a_0/a)$, we can rewrite the above two equations as
\begin{equation}\label{eff-eos-matter1}
H_m^2 = \frac{8 \pi G}{3c^2}\rho_m^0 e^{-3x(1+w_m^{\rm geo})} ,
\end{equation}
\begin{equation}\label{eff-eos-graviton1}
H_\delta^2 = \frac{8 \pi G}{3c^2}\rho_\delta^0
e^{-3x(1+w_\delta^{\rm geo})}
\end{equation}
Differentiating Eq. (\ref{eff-eos-graviton1}) with respect to $x$,
we obtain
\begin{equation}
2 H_\delta \frac{dH_\delta}{dx} = -3 (1 + w_\delta^{\rm geo}) H_\delta^2 .
\end{equation}
Rearranging this equation, one arrives at
\begin{equation}\label{eos-general1}
w_\delta^{\rm geo} = -1 - \frac{2}{3} \frac{1}{H_\delta} \frac{dH_\delta}{dx} = -1
-\frac{1}{3} \frac{d\ln H_\delta^2}{dx} \equiv w_{\rm DE},
\end{equation}
which confirms that $w_{\rm DE}$ differs from the native EoS
$w_g^{\rm nat}$ and equals to effective EoS $w_g^{\rm eff}$ when we
use Eq. (\ref{hubble}) as Hubble parameter. Note that our $w_{\rm
DE}$ is different from that (2.14) of Ref.~\cite{BPAL} even their
forms are the same because their Hubble equations are different. In
addition, we can obtain effective EoS for total cosmological fluid
as defined in Ref~\cite{AMA}:
\begin{equation}\label{eos-total}
w^{\rm eff} = -1 -\frac{1}{3}\frac{d \ln H^2}{dx} =
-1 + \frac{\Omega_m^0 \left ( \frac{a_0}{a}\right )^3 + \frac{\alpha_g^2}{6} a^2}
{\Omega_m^0 \left ( \frac{a_0}{a}\right )^3 + \frac{\alpha_g^2}{4} \left (1-a^2 \right )},
\end{equation}
which is depicted in Fig. 2. This EoS $w^{\rm eff}$ is exactly the
same as \begin{equation}
 w_{t}=\frac{p_t}{\rho_t}
\end{equation}
which is defined in (\ref{conserve-total}).

\section{Interaction mechanism}

We have revisited the Massive Graviton Dark Matter scenario (MGCDM)
which was recently considered as an alternative to dark energy
models. When introducing the native and effective equations of
state, it was shown that there is no phantom phase in the evolution
toward the far past of $a \to 0$ but the past accelerating phase
appears.

\begin{figure}[ht]
\label{a-fig} \mbox{\epsfxsize=8.5cm \epsffile{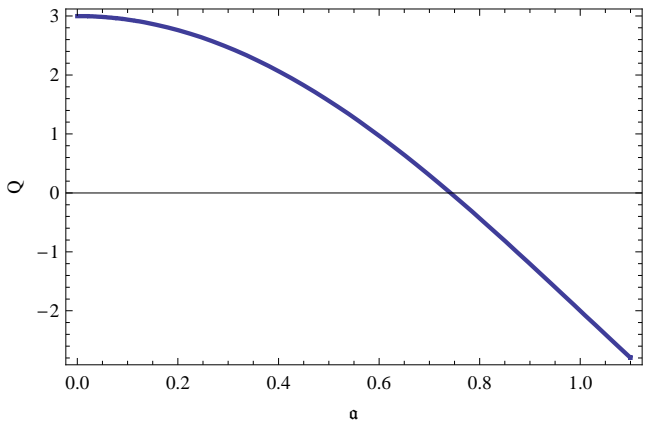}}
\caption{Time evolution for interaction  $Q$ as a function of scale
factor $a$ with 1 its coefficient. We note that  $Q=0$ at $a=0.74$.
}
\end{figure}
How do we understand the past accelerating phase in the  MGCDM? It
is well known that the interaction between two different  matters
give rises to acceleration~\cite{ZSBP,BPSZ, Myungs}.

In the MGCDM, the interaction between massive graviton and CDM
generates  acceleration in the past.  Graph of Fig. 5 shows that the
CDM($\rho_m,\Omega_m)$ decays to massive graviton $(\rho_g,
\Omega_g)$ for $Q>0$ and $0\le a < 0.74$, while massive graviton
decays to CDM  for $Q<0$ and $0.74 < a <1$. This picture is
consistent with the behavior of $(\Omega_g,\Omega_m)$ in Fig. 2. We
note that the effective EoS $w_g^{\rm eff}$ is an decreasing
function of $x$ in the backward direction  as is shown in Fig. 2. On
the other hand, it is noted  that the native EoS $w_{g}^{\rm nat}$
is meaningless because we are in the interacting picture.   The
graph of Fig. 6 shows the same graph depicted in terms of variable
$x$. It shows that the CDM decays to massive graviton for $Q>0$ and
$-\infty <  x < 0.98$, while massive graviton $(\rho_g, \Omega_g)$
decays to CDM ($\rho_m,\Omega_m)$ for $Q<0$ and $0.98 < x <1.16$.
The cosmic anti-fraction arisen from the CDM decay process to
massive graviton induces acceleration in the past
universe~\cite{ZSBP,BPSZ, Myungs}.
\begin{figure}[ht]
\label{qx-fig} \mbox{\epsfxsize=8.5cm \epsffile{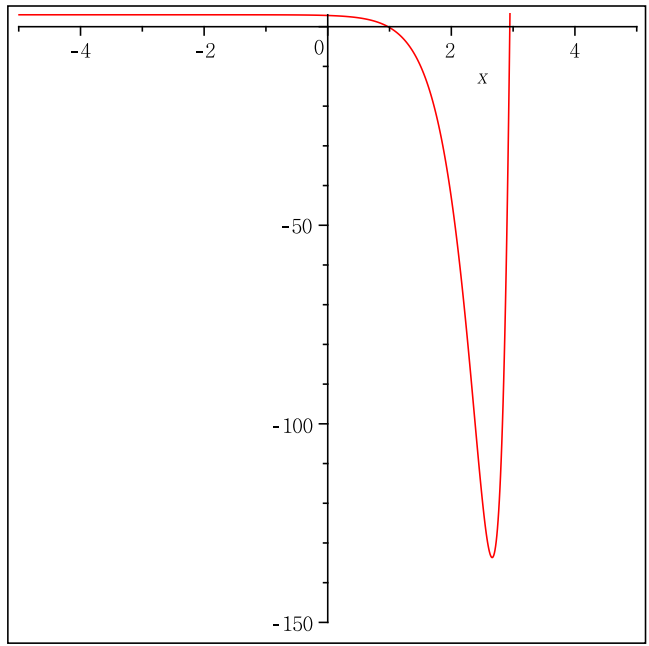}}
\caption{Time evolution for interaction  $Q$ as a function of $x$
with 1 its coefficient for $\Omega_m^0=0.27$, $\chi=3.1\times
10^{-4}$, and $\alpha_g=1.8$. We note that  $Q=0$ at $x=0.98$ and
2.95, and $x=1.16 (a=1)$. }
\end{figure}

Finally, we would like to mention that the condition of positive
energy density and negative pressure (\ref{ineq}) is not mandatory
to be fixed. In Appendix, we can extend this condition by
introducing the background static scaling factor $a_s$.

Consequently, any phantom phase des not appear in the evolution
toward the far past. The interaction between massive graviton and
CDM generates the past acceleration in the MGCDM.

\acknowledgments

This work was supported by the National Research Foundation of Korea
(NRF) grant funded by the Korea government (MEST) (No.2010-0028080).

\section*{Appendix: Background static  metric with scaling $a_s$}
If we choose the flat background metric  $g_0$ as
\begin{equation}\label{flat-metric-scale}
ds_0^2 = c^2 dt^2 - a_s^2 \left [ dr^2 + r^2 d\Omega_2^2 \right ],
\end{equation}
with $a_s$  an arbitrary scale factor,  the relevant quantities are
changed as
\begin{equation} \label{energy-massive-scale}
\rho_g = \frac{3 m_g^2 c^6 }{32 \pi G \hbar^2} (a_s^2 - a^2),
\end{equation}
\begin{equation} \label{pressure-massive-scale}
p_g = \frac{m_g^2 c^6 }{32 \pi G \hbar^2} a^2 (a^2-a_s^2).
\end{equation}
We note that the positive energy density and negative pressure is
required to explain the accelerating universe. In this case, one has
the inequality
\begin{equation} \label{mineq}
a^2 <a_s^2.
\end{equation}
Although the native EoS remains unchanged , the effective EoS
$w_g^{\rm eff}$ is changed as
\begin{equation}\label{eos-eff-graviton-scale}
w^{\rm eff}_g(a) = -
\frac{1-\frac{5}{3}\frac{a^2}{a_s^2}}{1-\frac{a^2}{a_s^2}},
\end{equation}
which is the same form as Eq. (\ref{eos-eff-graviton}) except
scaling factor $a^2_s$. Also,  the current scale factor is redefined
by
\begin{equation}\label{a0-scale}
a_0 = \sqrt{a_s^2- \frac{4}{\alpha_g^2}\Big(1 - \Omega_m^0 - \chi
\Omega_m^0 \Big)}.
\end{equation}
The condition of $\alpha_g$ is slightly changed as
\begin{equation}\label{low-limit-alpha-scale}
\alpha_g > \frac{2}{a_s} \sqrt{1-\Omega_m^0 - \chi \Omega_m^0},
\end{equation}
which  corresponds to the inequality condition on the mass of
massive graviton as
\begin{equation}\label{mass-low-scale}
m_g > \frac{2}{a_s} \sqrt{1-\Omega_m^0 - \chi \Omega_m^0} m_H .
\end{equation}
That is, by introducing the scaling $a_s$ of the background static
metric,  it is possible to  restrict the maximum scale factor for
the universe as in (\ref{mineq}).

\end{document}